\documentclass[12pt]{article}
\usepackage{epsfig}
\usepackage{amssymb}
\usepackage{amsmath,amsfonts,amssymb,graphicx}
\newcommand {\be}{\begin{equation}}
\newcommand {\ee}{\end {equation}}
\newcommand{\beq}{\begin{eqnarray}}
\newcommand{\eeq}{\end{eqnarray}}

\begin{document}
\vspace{-2cm}

\title{ Neutrino Oscillations With Recently Measured Sterile-Active Neutrino 
Mixing Angle}
\author{Leonard S. Kisslinger\\
Department of Physics, Carnegie Mellon University, Pittsburgh, PA 15213}
\maketitle
\date{}
\vspace{2mm}

\noindent
Keywords:Neutrino oscillations; Sterile neutrinos; New mixing angle

\noindent
PACS Indices:11.30.Er,14.60.Lm,13.15.+g
\vspace{0.25 in}
\begin{abstract}
  This  brief report is an extension of a prediction of neutrino 
oscillation with a sterile neutrino using parameters of the sterile neutrino
mass and mixing angle recently extracted from experiment.
\end{abstract}

\section{Introduction}

 Recently there was an investigation of how the neutrino transition probability
$\mathcal{P}(\nu_\mu \rightarrow \nu_e)$ is modified by the introduction of a 
sterile neutrino\cite{lsk14}, motivated in part by recent experiments on 
neutrino oscillations\cite{mini13}, which suggested the existence of at least 
one sterile neutrino.  The 3x3 U-matrix method of Sato and collaborators for 
three active neutrino oscillations\cite{as96,ks99} was extended to 4x4 with a
sterile neutrino. 

In the present brief report we use the formalism of Ref\cite{lsk14} with
parameters for the sterile neutrino that have recently been extracted by 
a global fit to neutrino oscillation data\cite{kopp13}.

\section{$\mathcal{P}(\nu_\mu \rightarrow \nu_e)$ Using a
4x4 U Matrix} 
 
Active neutrinos with
flavors $\nu_e,\nu_\mu,\nu_\tau$ and a sterile neutrino $\nu_s$ are 
related to neutrinos with definite mass by
\beq
\label{f-mrelation}
      \nu_f &=& U\nu_m \; ,
\eeq
where $U$ is a 4x4 matrix and $\nu_f,\nu_m$ are 4x1 column vectors, which
is an extension of the 3x3 matrix used in Refs.\cite{as96,ks99}.

As shown in Ref\cite{lsk14}, the transition probability $ \mathcal{P}(\nu_\mu 
 \rightarrow\nu_e)$, assuming $\delta_{CP}=0$ giving  $U^*_{ij}=U_{ij}$, is

\beq
\label{Pue}
 \mathcal{P}(\nu_\mu \rightarrow\nu_e) &=& U_{11}^2 U_{21}^2+
 U_{12}^2 U_{22}^2+ U_{13}^2 U_{23}^2+  \nonumber \\
  && U_{14}^2 U_{24}^2+ 2U_{11} U_{21} U_{12} U_{22} cos\delta L + \nonumber \\
  && 2(U_{11} U_{21} U_{13} U_{23}+ U_{12} U_{22} U_{13} U_{23})cos\Delta L+
\nonumber \\
  &&2U_{14}U_{24}(U_{11} U_{21}+U_{12} U_{22}+U_{13} U_{23})cos\gamma L \; ,
\eeq
with neutrino mass differences $\delta m_{ij}^2=m_i^2-m_j^2$, $\delta=\delta 
m_{12}^2/2E,\; \Delta=\delta m_{13}^2/2E,\; \gamma= \delta m_{j4}^2/2E$ (j=1,2,3),
where E is the energy and L the baseline. The neutrino mass differences are 
$\delta m_{12}^2=7.6 \times 10^{-5}(eV)^2$, $\delta m_{13}^2 = 2.4\times 10^{-3} 
(eV)^2$,  and from recent MiniBooNE analysis $\delta m_{j4}^2=0.9 (eV)^2$.
Note that in Ref\cite{lsk14} there is a typo in the third line of Eq(\ref{Pue}),
with $U_{13} U_{22}$ rather than the correct  $U_{13} U_{23}$.

Using  $c_{12}=.83,\;s_{12}=.56,\;s_{23}=c_{23}=.7071$, and $s_{13}=.15$,
(with $s_{ij}, c_{ij}=sin\theta_{ij},cos\theta_{ij}$),
\beq
\label{Uij}
   U_{11}&=& .822 c_\alpha \nonumber \\
   U_{12}&=&.554c_\alpha -.821 s_\alpha^2 \nonumber \\
   U_{13}&=&-.821s_\alpha^2c_\alpha-.554s_\alpha^2+.15c_\alpha  \nonumber \\
   U_{14}&=&.821s_\alpha c_\alpha^2+.554s_\alpha c_\alpha+.15s_\alpha  \\
   U_{21}&=& -.484c_\alpha  \nonumber \\
   U_{22}&=&.484s_\alpha^2 +.527 c_\alpha   \nonumber \\
   U_{23}&=&.699 c_\alpha-(-.484 s_\alpha c_\alpha+.527 s_\alpha)s_\alpha
\nonumber \\
   U_{24}&=&-.484 s_\alpha c_\alpha^2+.527s_\alpha c_\alpha+.699s_\alpha 
\nonumber \; ,
\eeq
with $\alpha$ the sterile-active neutrino mixing angle, $s_\alpha,c_\alpha$=
$sin(\alpha),cos(\alpha)$

A recent analysis of neutrino oscillation data\cite{kopp13} found 
$sin(\alpha)\simeq 0.16$. Using this, from
Eqs({\ref{Pue},{\ref{Uij}) we obtain the results shown in the figure. The
3x3 results with only active neutrinos are somewhat different than
those of Ref\cite{khj12} as the current value of $s_{13}=.15$ from the
Daya Bay experiment\cite{DB3-7-12,An2013} was not known at that time.

\clearpage
\begin{figure}[ht]
\begin{center}
\epsfig{file=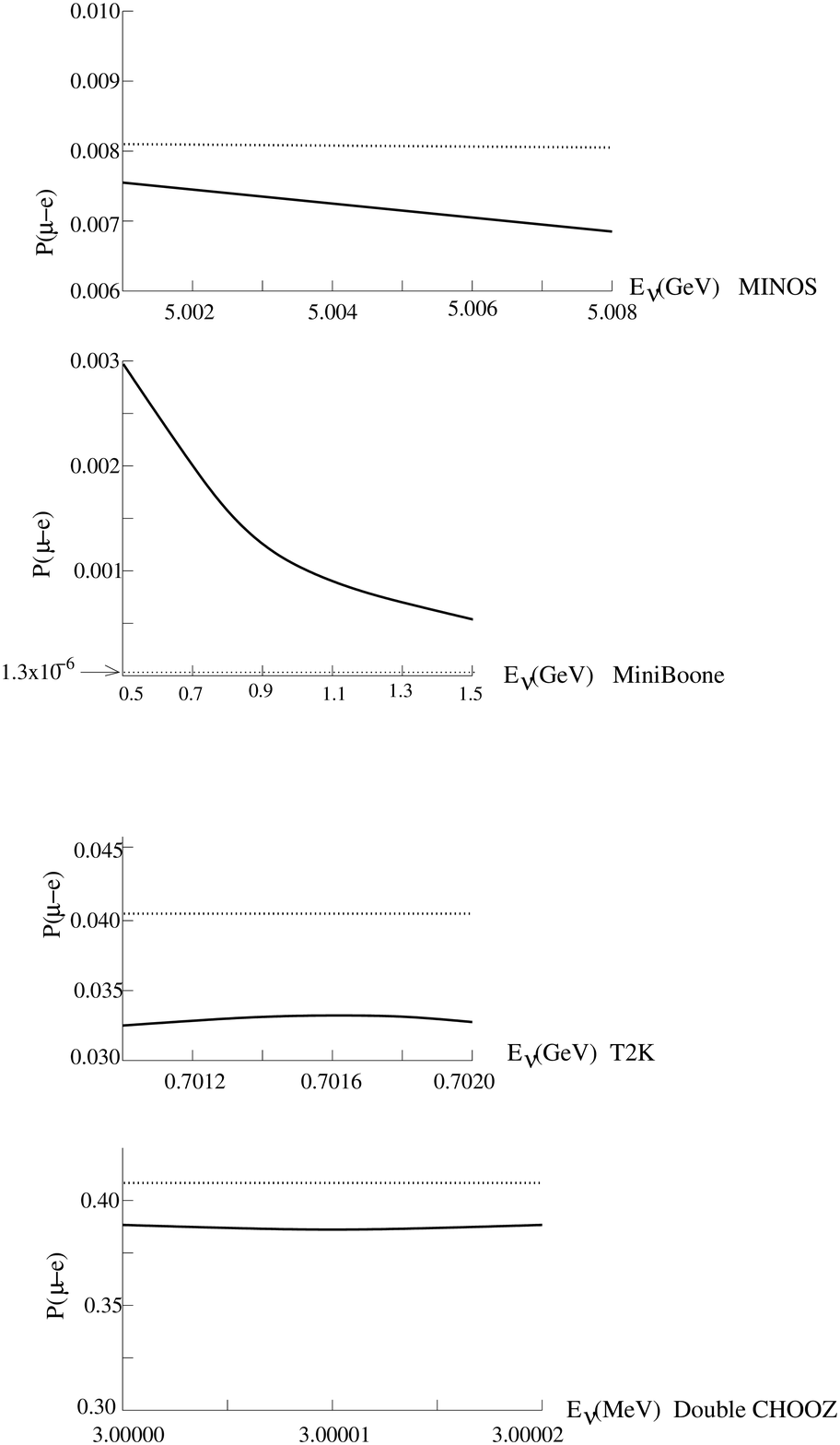,height=16cm,width=12cm}
\end{center}
\caption{\hspace{5mm} The ordinate is $\mathcal{P}(\nu_\mu \rightarrow\nu_e)$ 
for MINOS(L=735 km), MiniBooNE(L=500m), T2K(L=295 km), and Double 
CHOOZ(L=1.03 km) using the 4x4 U matrix with 
$\delta m_{j4}^2=0.9 (eV)^2$ and  $sin(\alpha)\simeq 0.16$. The dashed 
curves are for $\alpha=0$ (3x3).}
\end{figure}
\clearpage

\section{Conclusions}

As shown in the figure, the neutrino oscillation probability,
$\mathcal{P}(\nu_\mu \rightarrow \nu_e)$, is quite different for a model
with four neutrinos, $\nu_e,\nu_\mu,\nu_\tau,\nu_s$ than with three active
neutrinos. The effect of the sterile neutrino is largest for large
values of $E/L$, as seen in the figure for MiniBooNE, which has a baseline of
$L$=0.5 km. Although Double CHOOZ has a baseline of only 1.03 km, the energy 
is only about 3 MeV, so the effects of a sterile neutrino are smaller, but
are still important for the extraction of neutrino parameters from neutrino
oscillation data.

 In the present work we have used the sterile-active neutrino
mixing angle $\alpha$ ( $sin(\alpha)\simeq 0.16$), which is now known to be 
approximately correct for $\delta m_{j4}^2=0.9 (eV)^2$, rather than the mixing 
angles used in Ref\cite{lsk14}. Therefore the present results are more 
reliable for comparison with future neutrino oscillation experiments, 
inlcuding measurements of CP and T-reversal violations, than the previous 
publication.

\vspace{3mm}
\Large
{\bf Acknowledgements}
\vspace{3mm}

\normalsize
This work was carried out while the author was a visitor at Los Alamos
National Laboratory, Group P25, and the author thanks Dr. William C Louis III
for information on sterile neutrino parameters from MiniBooNE data and
several helpful discussions.

\end{document}